\documentclass[aps, amsmath, amssymb, 10pt, pra, twocolumn, longbibliography, superscriptaddress]{revtex4-1}

\usepackage{graphics,graphicx}
\usepackage{amsmath}
\usepackage{amsthm}
\usepackage{braket}
\usepackage{subfigure}
\usepackage{multirow}
\usepackage[colorlinks=true,urlcolor=blue,citecolor=blue,linkcolor=blue]{hyperref}
\usepackage{breakurl}

\newcommand{\beq}{\begin{equation}}
\newcommand{\eeq}{\end{equation}}

\theoremstyle{plain}

\theoremstyle{definition}

\begin{document}

%
%

\title{Experimental implementation of non-Clifford interleaved randomized benchmarking with a controlled-S gate}

\author{Shelly Garion}
\email[Corresponding author: ]{shelly@il.ibm.com}
\affiliation{IBM Quantum, IBM Research Haifa, Haifa University Campus, Mount Carmel, Haifa 31905, Israel}

\author{Naoki Kanazawa}
\email[Corresponding author: ]{knzwnao@jp.ibm.com}
\affiliation{IBM Quantum, IBM Research Tokyo, 19-21 Nihonbashi Hakozaki-cho, Chuo-ku, Tokyo, 103-8510, Japan}

\author{Haggai Landa}
\affiliation{IBM Quantum, IBM Research Haifa, Haifa University Campus, Mount Carmel, Haifa 31905, Israel}

\author{David C. McKay}
\affiliation{IBM Quantum, T.J.~Watson Research Center, Yorktown Heights, NY 10598, USA}

\author{Sarah Sheldon}
\affiliation{IBM Quantum, Almaden Research Center, San Jose, CA 9512, USA}

\author{Andrew W. Cross}
\affiliation{IBM Quantum, T.J.~Watson Research Center, Yorktown Heights, NY 10598, USA}

\author{Christopher J. Wood}
\affiliation{IBM Quantum, T.J.~Watson Research Center, Yorktown Heights, NY 10598, USA}

%
%

\begin{abstract}

Hardware efficient transpilation of quantum circuits to a quantum devices native gateset is essential for the execution of quantum algorithms on noisy quantum computers.
Typical quantum devices utilize a gateset with a single two-qubit Clifford entangling gate per pair of coupled qubits, however, in some applications access to a non-Clifford two-qubit gate can result in more optimal circuit decompositions and also allows more flexibility in optimizing over noise.
We demonstrate calibration of a low error non-Clifford Controlled-$\frac{\pi}{2}$ phase (CS) gate on a cloud based IBM Quantum computing using the Qiskit Pulse framework.
To measure the gate error of the calibrated CS gate we perform non-Clifford CNOT-Dihedral interleaved randomized benchmarking.
We are able to obtain a gate error of $5.9(7) \times 10^{-3}$ at a gate length 263 ns, which is close to the coherence limit of the associated qubits, and lower error than the backends standard calibrated CNOT gate.

\end{abstract}

\maketitle

%
%

\section{Introduction}

Quantum computation holds great promise for speeding up certain classes of problems, however near-term applications are heavily restricted by the errors that occur on present day noisy quantum devices~\cite{Preskill2018}.
To run a computation on a quantum processor requires first calibrating a universal gate set -- a small set of gates which can be used to implement an arbitrary quantum circuit -- which has low error rates, and then transpiling the circuit to this set of gates.
This transpilation should be done in a hardware-efficient manner to reduce the overall error by minimizing the use of the highest error gates \cite{Corcoles2019}.
Two of the most significant error sources on current devices are incoherent errors due to interactions with the environment, quantified by the coherence times of device qubits, and calibration errors in the gates used to implement a quantum computation~\cite{IterativeRB, RBPurity2}.

If a gate set could be perfectly calibrated the coherence time of the qubits would set the fundamental limit on error rates without active error correction.
Thus the goal of gate calibration is to get as close to the coherence limit as possible.
Current quantum hardware typically use a gate set consisting of arbitrary single-qubit rotations and a single entangling two-qubit gate \cite{Mckay2018}.
State of the art single-qubit gate error rates in these systems approach $2 \times 10^{-4}$~\cite{Zgates},
where two-qubit gate errors are around $10^{-3}$~\cite{Gambetta2017, Xu2020, Foxen2020}, see also Appendix \ref{sec:device_info}.
In superconducting qubit systems using fixed-frequency transmon qubits a microwave-only two-qubit entangling gate may be implemented using the cross-resonance (CR) interaction~\cite{Chow2011}.
The CR interaction can be used to implement a high fidelity Controlled-NOT (CNOT) gate~\cite{Sheldon2016}.
 Gate sets with a Clifford two-qubit like CNOT are appealing as a variety of averaged errors in a Clifford gateset can be can be robustly measured using various randomized benchmarking (RB) protocols~\cite{RB10,RBint,RBleak,McKay2016,RBPurity2,RBcorr}.

In some cases it may be favorable to introduce an additional two-qubit gate to a gate set if it enables more hardware efficient compilation of relevant circuits, however this adds the overhead of additional calibration and characterization of the gate errors.
One such gate is the Controlled-Phase (CS) gate, which is a non-Clifford two-qubit entangling gate that is universal when combined with the Clifford group~\cite{NonCliff}.
The CS gate is particularly attractive to fixed-frequency transmon qubit systems as it can be implemented using the CR interaction, since it is locally equivalent to $\sqrt{\text{CNOT}}$.
This means it can be calibrated using the same techniques as the CNOT gate, but with a shorter gate duration or lower power, potentially leading to a higher fidelity two-qubit gate when calibrated close to the coherence limit.
Furthermore the CS gate is a member the CNOT-Dihedral group and can be benchmarked using CNOT-Dihedral randomized benchmarking~\cite{NonCliff}.
Recently an optimal decomposition algorithm for two-qubit circuits into the Clifford + CS gates was developed~\cite{CliffCS}.
This method minimizes the number of non-Clifford (CS) gates, which is important in the context of quantum error correction as non-Clifford gates require additional resources such as magic-state distillation to prepare fault-tolerantly \cite{Bravyi2005}.
However, in non-fault tolerant near term devices it is often preferable to minimize the total number of two-qubit gates in a decomposition rather than non-Clifford gates.
An optimal decomposition for gates generated by the CNOT-Dihedral in terms of the number of CNOT and CS gates has also recently been developed~\cite{Garion2020}.
Another example is the Toffoli gate which can be decomposed into 6 CNOT gates and single qubit gates, but requires only 5 two-qubit gates in its decomposition if the CS and $\text{CS}^{-1}$ gates are also available~\cite{Toffoli}.

In this work we calibrate CS and $\text{CS}^{-1}$ gates of varying durations on an IBM Quantum system and benchmark the gate error rates by performing the first experimental demonstration of interleaved CNOT-Dihedral randomized benchmarking.
For specific gate durations we are able to obtain a high-fidelity CS gate approaching the coherence limit, which due to the shorter CR interaction time results in a lower error rate than can be obtained for a CNOT gate.
In addition to RB we also compute the average gate error of the CS gate using two-qubit quantum process tomography (QPT) and compare to the values obtained from RB.
Pulse-level calibration was done using Qiskit Pulse~\cite{Alexander2020}, and the RB and QPT experiments were implemented using the open source Qiskit computing software stack~\cite{Qiskit} through the IBM Quantum cloud provider.

%
%

\section{CNOT-Dihedral Randomized Benchmarking}
\label{sec:rb_protocol}

We describe the protocol for estimating the average gate error of the CS gate using interleaved CNOT-Dihedral Randomized Benchmarking, which is a natural generalization of the CNOT-Dihedral RB procedure described in~\cite{NonCliff} with interleaved RB~\cite{RBint} to estimate individual gate fidelities for the CS gate
$$CS = \begin{pmatrix} 1 & 0 & 0 & 0 \\ 0 & 1 & 0 & 0 \\
    0 & 0 & 1 & 0 \\ 0 & 0 & 0 & i \end{pmatrix}.$$

In the following we let $G$ denote the CNOT-Dihedral group on $n$ qubits and $g \in G$ denote a unitary element of $G$.
Here the CNOT-Dihedral group is generated by the single-qubit gates $X, T$ and the CNOT gate.
More precisely,
$$G =  \langle X_i, T_i, {\rm CNOT}_{i,j} \rangle / \langle \lambda I: \lambda \in \mathbb{C} \rangle,$$
where $i, j \in \{0,\dots,n-1\} , i \neq j$.
We denote $I$, $X$, $Y$, $Z$ as the single-qubit Pauli matrices and $T =\begin{pmatrix} 1 & 0 \\0 & e^{\pi i / 4} \end{pmatrix}$.

\subsection*{Step 1: Standard CNOT-Dihedral benchmarking.}

Randomly sample $l$ elements $g_{j_1},\dots,g_{j_l}$ uniformly from $G$, and compute the ($l+1$)th element from the inverse of their composition,
$g_{j_{(l+1)}} = (g_{j_l} \circ \dots \circ g_{j_1})^{-1}$.
Denote by ${\bf j}_l$ the $l$-tuple $(j_1,\dots,j_l)$.
For each sequence, we prepare an input state $\rho$, and apply the composition of the $l+1$ gates that ideally would be
$$S_{{\bf j}_l} := g_{j_{(l+1)}} \circ g_{j_l} \circ \dots \circ g_{j_1},$$
and then measure the expectation value of an observable $E$.

Assuming each gate $g_i$ has an associated error $\Lambda_i(\rho)$, the sequence $S_{{\bf j}_l}$ is implemented as
\begin{equation}
\tilde{S}_{{\bf j}_l} := \Lambda_{j_{(l+1)}} \circ g_{j_{(l+1)}} \circ
\bigl( \bigcirc_{i=1}^l [\Lambda_{j_i} \circ g_{j_i}] \bigr)
\end{equation}

The expectation value of $E$ is $\langle E\rangle_{{\bf j}_l} = Tr[E \tilde{S}_{{\bf j}_l}(\rho)]$.
Averaging this overlap over $K$ independent sequences of length $l$ gives an estimate of the average sequence fidelity
\begin{equation}
F_{seq}(l,E,\rho) := Tr[E \tilde{S}_l(\rho)]
\end{equation}
where $\tilde{S}_l(\rho) := \frac{1}{K} \sum_{{\bf j}_l} \tilde{S}_{{\bf j}_l} (\rho)$ is the average quantum channel.

We decompose the input state and this final measurement operator in the Pauli basis $\mathcal{P}$ (an orthonormal basis of the  $n$-qubit Hermitian operators space, constructed of single-qubit Pauli matrices).
This gives
$\rho = \Sigma_P x_P P / 2^n$ and $E' = \Sigma_P e_P P$.
Given that the gate errors are close to the average of all errors \cite{NonCliff}, the average sequence fidelity is
$$F_{seq}(l,E,\rho) = A_Z \alpha_Z^l + A_R \alpha_R^l + e_I$$
where $A_Z = \Sigma_{P \in \mathcal{Z} \setminus \{I\}} e_P x_P$ and $A_R = \Sigma_{P \in \mathcal{P} \setminus \mathcal{Z}} e_P x_P$, with $ \mathcal{Z}$ being tensor products of $Z$ and $I$ gates.

Each of the two exponential decays $\alpha_Z^l$ and $\alpha_R^l$ can be observed by choosing appropriate input states.
For example, if we choose the input state $|0 \dots 0 \rangle$ then $F_{seq} = e_I + A_0 \alpha_Z^l$ where $A_0 = \Sigma_{P \in \mathcal{Z} \setminus \{I\}} e_P$.
On the other hand, if we choose $|+ \dots + \rangle$ then $F_{seq} = e_I + A_+ \alpha_R^l$ where $A_+ = \Sigma_{P \in \mathcal{X} \setminus \{I\}} e_P$, with $ \mathcal{X}$ tensor products of $X$ and $I$ gates.

The channel parameters $\alpha_Z$ and $\alpha_R$ can be extracted by fitting the average sequence fidelity to an exponential.
From $\alpha_Z, \alpha_R$ the average \emph{depolarizing channel parameter} $\alpha$ for a group element $g$ is given by
\begin{equation}\label{eq:alpha}
\alpha = (\alpha_Z + 2^n \alpha_R)/(2^n+1)
\end{equation}
and the corresponding \emph{average gate error} is given by
\begin{equation}\label{eq:epc}
r = (2^n-1)(1-\alpha)/2^n.
\end{equation}

\subsection*{Step 2: Interleaved CNOT-Dihedral sequences.}

Choose a sequence of unitary gates where the first element $g_{j_1}$ is chosen uniformly at random from $G$, the second is always chosen to be $g$, and alternate between uniformly random elements from $G$ and fixed $g$ up to the $l$-th random gate.
The $(l+1)$ element is chosen to be the inverse of the composition of the first $l$ random gates and $l$ interlaced $g$ gates, $g_{j_{(l+1)}} = (g \circ g_{j_l} \circ \dots \circ g \circ g_{j_1})^{-1}$.
We adopt the convention of defining the length of a sequence by the number of random gates $l$.

For each sequence, we prepare an input state $\rho$, apply
$$\nu_{{\bf j}_l} := g_{j_{(l+1)}} \circ g \circ g_{j_l} \circ \dots \circ g \circ g_{j_1}$$
and measure an operator $E$.

Assuming that the gate $g$ has an associated error $\Lambda_g(\rho)$ and that each gate $g_i$ has an associated error $\Lambda_i(\rho)$, the sequence $\nu_{{\bf j}_l}$ is implemented as
\begin{equation}\label{eq:int}
\tilde{\nu}_{{\bf j}_l} := \Lambda_{j_{(l+1)}} \circ g_{j_{(l+1)}} \circ
\bigl( \bigcirc_{i=1}^l [\Lambda_g \circ g \circ \Lambda_{j_i} \circ g_{j_i}] \bigr).
\end{equation}

The overlap with $E$ is $Tr[E \tilde{\nu}_{{\bf j}_l}(\rho)]$.
Averaging this overlap over $K$ independent sequences of length $l$ gives an estimate of the new sequence fidelity
$$F_{\overline{seq}}(l,E,\rho) := Tr[E \tilde{\nu}_l(\rho)]$$
where $\tilde{\nu}_l(\rho) := \frac{1}{K} \sum_{{\bf j}_l} \tilde{\nu}_{{\bf j}_l} (\rho)$ is the average quantum channel.

Similarly to Step 1, we fit $F_{\overline{seq}}(l,E,\rho)$ and obtain the depolarizing parameter $\alpha_{\bar g}$, according to Eq.~(\ref{eq:alpha}).
Using the values obtained for $\alpha$
and $\alpha_{\bar g}$
, the gate error of $\Lambda_g$, which is  given by
\begin{equation} \label{eq:average_gate_error}
r_g^{\rm rb} = \frac{(2^n-1)(1-\alpha_{\bar g}/\alpha)}{2^n},
\end{equation}
and must lie in the range $[r_g^{\rm rb}-\epsilon, \min(r_g^{\rm rb}+\epsilon, 1)]$, where $\epsilon$ can be estimated using~\cite{RBint} Eq.~(5), or \cite{Kimmel_2014} Eq. (VI.1).
Note that one has to be careful in interpreting the results of an interleaved experiment, as in some cases $\epsilon$ might be large compared to $r_g^{\rm rb}$.

%
%

\section{Implementing the Controlled-{\it S} gate} \label{sec:cs_calibration}

We calibrate CS gates of varying gate durations using Qiskit Pulse and measure the average gate error using the interleaved CNOT-Dihedral RB protocol in \ref{sec:rb_protocol}.
We use the CR pulse sequence as a generator of two-qubit entanglement \cite{Rigetti2010, Chow2011}.
The CR pulse is realized by irradiating one (control) qubit with a microwave pulse at the transition frequency of another (target) qubit.
The stimulus drives the quantum state of the target qubit with the direction of rotation depending on the quantum state of the control qubit.
This controlled rotation is used to create two-qubit entangling gates such as CNOT and CS.

The two-qubit system driven by the CR pulse with amplitude $A$ and phase $\phi$ can be approximated by an effective block-diagonal time-independent Hamiltonian \cite{Magesan2018, Malekakhlagh2020}
\begin{align} \label{eq:crham}
  \overline{H}_{\text{CR}}(A, \phi) =& \sum_{P = I, X, Y, Z} \frac{\omega_{ZP}(A, \phi)}{2} Z \otimes P \\ \notag
  & + \sum_{Q = X, Y, Z} \frac{\omega_{IQ}(A, \phi)}{2} I \otimes Q,
\end{align}
where the qubit ordering is control $\otimes$ target, and  $\omega_{ZP}$ and $\omega_{IQ}$ represent the interaction strength of the corresponding Pauli Hamiltonian terms.
In the absence of noise, the ideal CR evolution for a constant-amplitude pulse is written as an unitary operator
\begin{align} \label{eq:cr_evolution}
  U_{CR}(A, \phi)= \exp\left\{- i t_{CR} \overline{H}_{CR}(A, \phi)\right\},
\end{align}
where $t_{CR}$ is the length of the CR pulse.
We also define the unitary operator created by an arbitrary two-qubit generator as
\begin{equation}
[BC]_\theta = \exp\left\{-i\frac{\theta}{2} (B\otimes C)\right\}
\end{equation}
where $B$, $C$ are arbitrary single qubit operators, and we use $[BC] \equiv [BC]_\pi$.

As can be seen by examining Eq.~\eqref{eq:crham}, the CR pulse induces three entangling interaction terms ($ZX$, $ZY$, and $ZZ$), in addition to potentially many unwanted local rotations with different amplitudes.
By appropriately calibrating the phase of the CR drive $\phi$, the $ZX$ term is the dominant term among the interactions and is the key term for executing two-qubit gates in this system.
As with the standard CNOT gate, we can compose a CS gate by isolating the ZX interaction with a refocusing sequence and single qubit pre- and post-rotations:
\begin{align} \label{eq:csgate}
  CS = [IH] \circ [IX]_{\frac{\pi}{4}} \circ [ZI]_{\frac{\pi}{4}} \circ [ZX]_{-\frac{\pi}{4}} \circ [IH],
\end{align}
where $H$ is the Hadamard operator.
As shown in Eq.~\eqref{eq:csgate}, we need to develop the calibration procedure to find an amplitude $A$ and a phase $\phi$ where $|\omega_{ZX}|t_{CR} = \pi/4$ and the other terms become zero.
The CR Hamiltonian includes a large $ZI$ term as a result of the off-resonant driving of the control qubit; $IX$, $ZZ$ and $IZ$ can also be large for transmon qubits \cite{Magesan2018}.
However, the strengths of $ZZ$ and $IZ$ terms are expected to be negligibly weak in our device.
We note that both $ZI$ and $IX$ terms commute with the $ZX$ term of interest, while $ZI$ and $ZX$ terms anti-commute with the inversion of the control qubit $XI$.
In addition, the $ZI$ term is the even function and both $IX$ and $ZX$ terms are odd functions of the drive amplitude $A$.
Accordingly, we can effectively eliminate the impact of those unwanted terms with the two-pulse echoed CR sequence \cite{Corcoles2013} expressed as
\beq U_\text{echo}(A, \phi) =  [XI] \circ U_{CR}(-A, \phi) \circ [XI] \circ U_{CR}(A, \phi).\label{eq:echo}\eeq
This sequence consists of two CR pulses with opposite drive amplitude, each one followed by a $\pi$-rotation refocus pulse $XI$ on the control qubit.
Here we also assume the negligible impact of the $IY$ term which is generally introduced by the physical crosstalk between the control and the target qubit \cite{Sheldon2016}.

\subsection{Gate Calibration and Benchmarks}

To experimentally implement the CS and $\text{CS}^\dag$ gates we use the 27 qubit IBM Quantum system \texttt{ibmq\_paris} with fixed-frequency and dispersively coupled transmon qubits.
Qubit 0 and the qubit 1 of this system are assigned as the control and the target qubit, respectively.
The resonance frequency and anharmonicity of the control (target) qubit are 5.072 (5.020) GHz and -336.0 (-321.0) MHz.

The pulses realized in practice are not constant-amplitude pulses, rather the amplitude is increased and decreased smoothly.
We implement the CR pulse as a flat top Gaussian, with flat-top length $\tau_\text{sq}$, and Gaussian rising and falling edges each with length $\tau_\text{edge}$ ($\tau_{\text{CR}} = \tau_\text{sq} + 2\tau_\text{edge}$).
We use a constant Gaussian edge with $\tau_\text{edge} = 28.16$ ns with $14.08$ ns standard deviation and vary the length of the duration of the square flat-top pulse $\tau_\text{sq}$.
The minimum pulse duration is $\tau_\text{sq} = 0$ ns, yielding a pure Gaussian shape.
The overhead of single-qubit gates in the echoed CS sequence in Eq.~\eqref{eq:csgate} for the \texttt{ibmq\_paris} backend is 106.7 ns, giving a total echoed CS gate time of $\tau_{\text{CS}} = 2 \tau_{\text{CR}} + 106.7 ~\text{ns}$.
The single-qubit gates are optimized by merging consecutive rotations using the Qiskit circuit transpiler with \texttt{optimization\_level} = 1 followed by conversion to a pulse schedule \cite{Alexander2020}.

\begin{figure}[tb!]
  \includegraphics[clip,width=0.95\linewidth]{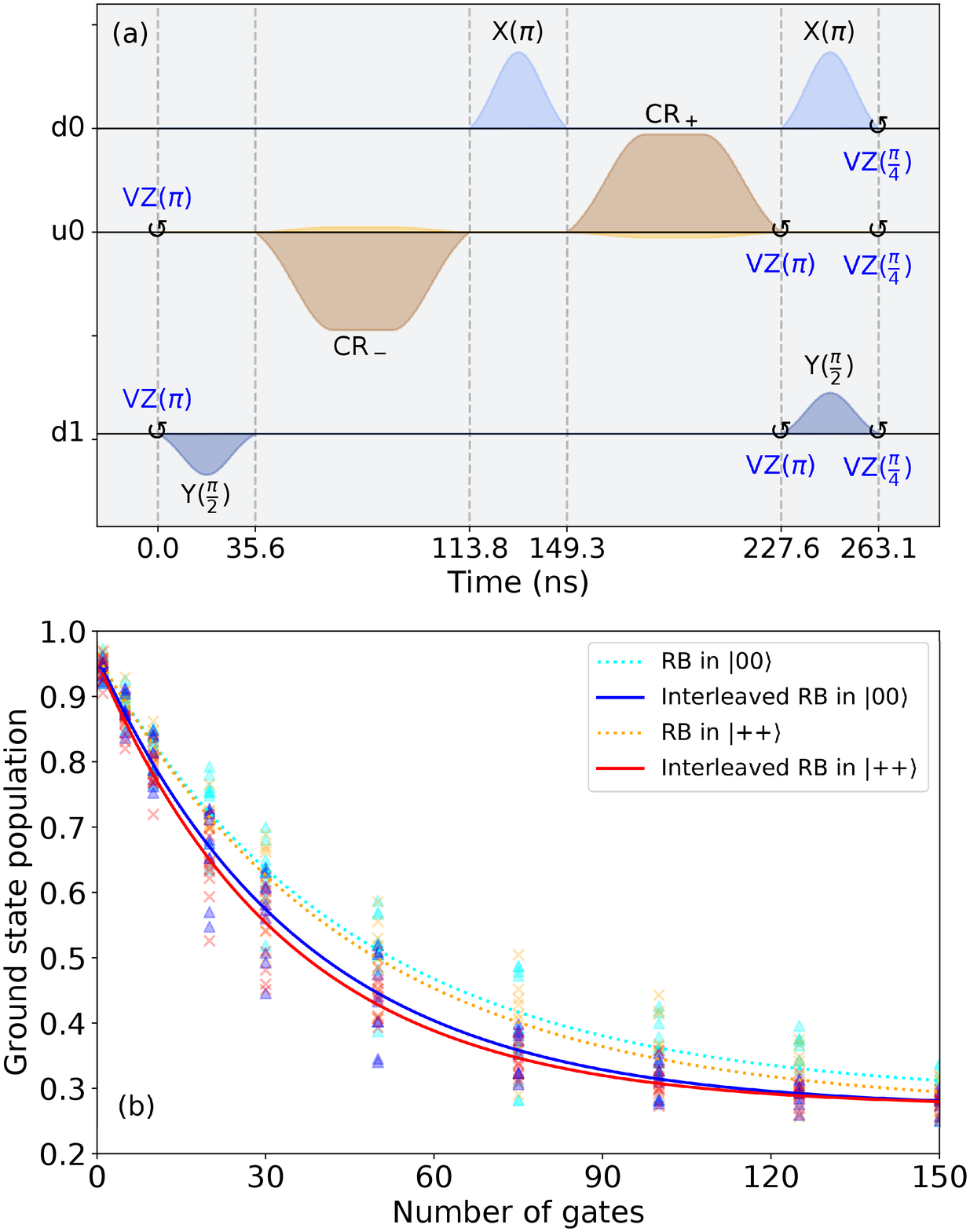}
  \caption[justification=centerlast]{
  The CS gate realized with a closed-loop calibration.
  (a) Pulse schedule with the flat-top width $\tau_{\rm sq}$ = 21.3 ns.
  The schedule consists of two CR pulses $CR_{-}$ and $CR_{+}$ on the \texttt{ControlChannel} \texttt{u0} with echo pulses $X(\pi)$ applied on \texttt{DriveChannel} \texttt{d0} of the control qubit.
  Local gates in Eq. \eqref{eq:csgate} are also applied to the \texttt{DriveChannel} \texttt{d1} of the target qubit.
  Pulse instructions in \texttt{d0} and \texttt{d1} are played in the rotating frame of the control and the target qubits, respectively.
  The \texttt{ControlChannel u0} is physically connected to the control qubit, whereas pulses are played in the rotating frame of the target qubit to drive CR interaction.
  A Circular arrow of \texttt{VZ}($\theta$) represents the virtual-Z rotations with rotation angle $\theta$.
  (b) CNOT-Dihedral interleaved RB. Dotted lines show fit curves of the ground state population measured by standard RBs in $|00\rangle$ and $|++\rangle$ basis, while solid lines show fits of interleaved RB.
  Triangle and cross symbols show raw experiment data of 10 different random circuits.
  }
  \label{fig:cs_calibration_results}
\end{figure}

We performed calibration to a CR rotation angle $\omega_{ZX}(A, \phi) \tau_\text{CR} \simeq \pi/4$ for different values of $\tau_\text{sq}$.
This was done by first performing a rough calibration of $(A, \phi)$ by scanning those parameters, followed by the closed-loop fine calibration with standard error amplification sequences (see Appendix \ref{sec:gate_calibration} for details).
The calibrated pulse schedule of the CS gate with $\tau_{\rm sq} = 21.3$ ns ($\tau_{CS} = 263.1$ ns) is shown in Fig. \ref{fig:cs_calibration_results}(a).

The average gate error of the calibrated CS gate is evaluated by using the interleaved CNOT-Dihedral RB with 10 sequence lengths $l \in (1, 5, 10, 20, 30, 50, 75, 100, 125, 150)$, and 10 samples for each $l$.
Each experiment is executed 1024 times for both input states $\ket{00}$ and $\ket{++}$ both with and without interleaving the CS gate.
An example of measured RB decay curves for $\tau_{\rm sq}$ = 21.3 ns are shown in Fig. \ref{fig:cs_calibration_results}(b).
The exponential fit of the decay curves yields $\alpha = 9.78(1) \times 10^{-1}$ and $\alpha_{\bar{g}} = 9.73(1) \times 10^{-1}$, giving an estimated average gate error of the CS gate of $r_g^{\rm rb} = 5.2(7) \times 10^{-3}$.
In addition to RB we also perform quantum process tomography (QPT)~\cite{Mohseni2008} and compute the average gate fidelity from the reconstructed process, see the Appendix \ref{sec:qpt_experiment} for the details.
The average gate error calculated from the tomographic fit for $\tau_{\rm sq} = 21.3$ ns was $r_g^{\rm qpt} = 1.36 \times 10^{-2}$ which is slightly higher but still comparable to the value estimated from the interleaved CNOT-Dihedral RB experiment.

\subsection{Gate Duration Dependence}

\begin{figure}[t!]
  \includegraphics[width=0.95\linewidth]{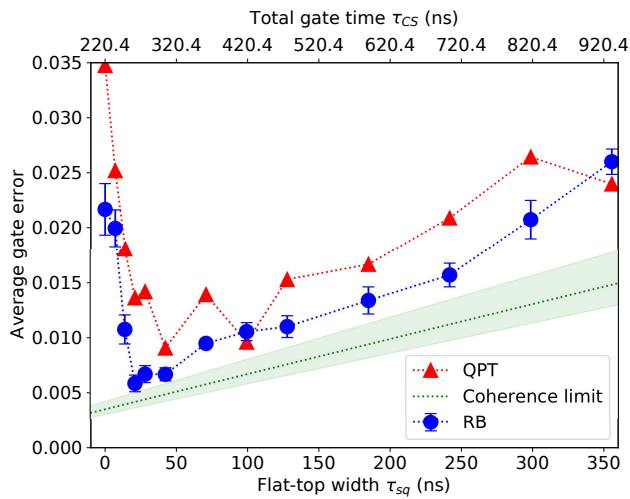}
  \caption[justified]{
  Average gate errors as a function of the flat-top width of the CR pulse $\tau_\text{sq}$ estimated by different benchmark techniques.
  The corresponding total gate time $\tau_{CS}$ is shown in the top axis.
  Blue circles and red triangles represent $r_g^{\rm rb}$ and $r_g^{\rm qpt}$, respectively.
  The Green dotted line shows the theoretical lower bound of the average gate error calculated by the total gate time $\tau_{CS}$ and the average $T_1$ and $T_2$ values of the qubits during the experiment.
  The filled area represents the coherence limit with $T_1$ and $T_2$ values with variance of 1$\sigma$.
  See text for a detailed discussion.
  }
  \label{fig:gate_speed_dependency}
\end{figure}

We perform the same calibration and benchmarking procedures for different flat-top width $\tau_{\rm sq}$ from 0 ns to 355.6 ns ($\tau_{CS}$ from 219.3 ns to 930.5 ns) and measure the average gate errors by both the interleaved CNOT-Dihedral RB experiment and QPT.
In this experiment, we use a reduced set of RB sequence lengths $l \in (1, 10, 25, 50, 100, 150)$ to reduce the total number of experiments while keeping the accuracy of the estimated gate error high.

We measure the qubit coherence times $T_1$ and $T_2$ with relaxation and Hahn echo sequences \cite{Bylander2011}, respectively, to monitor the stability of physical properties of qubits.
These experiments are inserted immediately before each calibration experiment and yield coherence times of $T_1 = 59.6 \pm 15.6$ ($77.1 \pm 7.3$)$\,\mu$s and $T_2 = 92.5 \pm 22.1$ ($69.1 \pm 4.8$)$\,\mu$s for the control (target) qubit during the experiment.
Here, the error bars correspond to the standard deviation over the duration of the whole set of calibration and benchmarking experiments.
A lower bound of gate error at $\tau_{\rm sq}$ is calculated based on the measured $T_1$ and $T_2$ values with the total gate duration $\tau_{\text{CS}}$, see Appendix G of Ref. \cite{Sundaresan2020}.
The \texttt{coherence\_limit} function in Qiskit Ignis \cite{qiskit-ignis} is used for the calculation, presented in Fig.~\ref{fig:gate_speed_dependency}.
The device was accessed via the cloud through a fair-share queuing model used in IBM Quantum systems.
The time in between experiments was about 168 minutes on average, thus the experiment could be subject to some parameter fluctuations due to noise with a long characteristic time \cite{Schlor2019}.

Nevertheless, as $r_g^{\rm rb}$ in Fig.~\ref{fig:gate_speed_dependency} shows, our calibration method provides highly accurate results and allows to approach the coherence limit for appropriately chosen gate times.
This dependence on $\tau_\text{sq}$ agrees well with the slope predicted by the coherence limit for $\tau_{\rm sq} \gtrsim 21.3$ ns.
We also plot $r_g^{\rm qpt}$ as a reference since QPT is conventionally used to evaluate the performance of non-Clifford gates.
These lines show reasonable agreement though $r_g^{\rm qpt}$ tends to show slightly higher gate errors than $r_g^{\rm rb}$.
This is expected as QPT is sensitive to state preparation and measurement errors, though measurement errors have been reduced by using readout error mitigation.
The interleaved CNOT-Dihedral RB experiment requires only 24 circuit executions per single error measurement, while the two-qubit QPT requires 148 circuit executions with the readout error mitigation.
The smaller experimental cost to measure $r_g^{\rm rb}$ enables us to average the result over 10 different random circuits, which is empirically sufficient to obtain a reproducible outcome, at a practical queuing time with \texttt{ibmq\_paris}.
The nearly stable offset of $r_g^{\rm rb}$ from the coherence limit possibly indicates the presence of coherent errors due to imperfection of calibration.

In the region $\tau_{\rm sq} \lesssim 21.3$ ns, both gate errors show a significant increase from the coherence limit.
In this regime the drive amplitude of the CR pulse rapidly increases in order to guarantee that the total accumulated rotation angle is $\pi/4$ for shorter $\tau_{\rm CR}$.
The amplitude of crosstalk $\sqrt{\omega_{IX}^2 + \omega_{IY}^2}$ measured at $\tau_\text{sq} = 0$ ns is 176.2 kHz, while one at $\tau_\text{sq} = 355.6$ ns is 19.4 kHz.
Although the $IX$ term is refocused and has negligible contribution, the remained $IY$ term can still impact on the measured gate errors.
Thus, at $\tau_\text{sq} = 0$ we calibrate a CS gate with a compensation tone on the target qubit to suppress the physical crosstalk between qubits (see Appendix \ref{sec:xtalk_estimation} for details).
The calibrated pulse sequences with and without the compensation tone yield $r_g^{\rm rb}$ of $2.1(3) \times 10^{-2}$ and $2.2(2) \times 10^{-2}$, respectively.
These comparable results indicate the physical crosstalk is relatively suppressed in this quantum device and other noise sources are dominant for $\tau_{\rm sq} \lesssim 21.3$ ns.
For example, at high power the perturbation theory used to obtain the average CR Hamiltonian may break down, and hence also calibration scheme based on this decomposition.

The reasons for imperfection of two-qubit gates in superconducting qubits have been investigated and associated with various mechanisms such as nonideal signal generation, residual $ZZ$ coupling, CR-induced $ZZ$ interaction \cite{Ganzhorn2020, Noguchi2020, Ku2020}, and leakage to the higher energy levels \cite{McKay2016, Rol2019}.
Although a further analysis of the error mechanisms in this regime of high-power pulses is beyond the scope of this study, initial results indicate that coherent population transfer out of the two-qubit manifold into the higher levels, and $ZZ$ interaction terms, are not the relevant mechanisms \cite{KanazawaUnpublished}.
At the same time, the coherence limit can be further lowered by reducing the time spent on single-qubit gates.
At $\tau_{\rm sq} = 21.3$ ns with the minimum $r_g^{\rm rb}$ of $5.9(7) \times 10^{-3}$, the refocusing pulse and local rotations occupy 40\% of the total gate time $\tau_{CS}$, yielding a non-negligible impact on the gate error.

The interleaved CNOT-Dihedral Randomized Benchmarking technique can be used to evaluate any quantum gate in the CNOT-Dihedral group regardless of its physical qubit implementation. The calibration protocol is also general to devices which are capable of driving the CR interaction.

%
%

\section{Conclusion} \label{sec:conclusion}

We have demonstrated calibration of a high fidelity non-Clifford CS gate on 27 qubit IBM Quantum system \texttt{ibmq\_paris}.
This gate is not currently included in the standard basis gates of IBM Quantum systems, and it was calibrated and benchmarked entirely using open source software available in Qiskit.
Since the CS gate is non-Clifford, robust characterization of the average gate error cannot be done using standard RB.
To benchmark performance of the non-Clifford gate we performed the first experimental demonstration of two-qubit interleaved CNOT-Dihedral RB, which allow efficient and robust characterization of a universal gateset containing the CS gate.

We obtained a minimal gate error of $5.9(7) \times 10^{-3}$ with appropriately shaped echoes and a total gate time of $263.1$ ns.
The gate error reported for the standard two-qubit CNOT gate provided by \texttt{ibmq\_paris} is $1.3 \times 10^{-2}$. Thus the presented CS gate error is comparable with half the CNOT error.
By performing RB and QPT for a variety of gate lengths we were also able to study the performance of the CS gate in different regimes and observed a break down in performance if gate lengths were reduced below the best value obtained for $263.1$ ns.
This is consistent with previous literature on CNOT calibration using the cross-resonance interaction in the high power regime.

The expansion of the native two-qubit gateset of a Cloud quantum device with additional low error calibrated gates allows for improved hardware efficient transpilation of quantum circuits.
This is important for executing quantum algorithms on noisy quantum devices without error correction, and for reducing the error correction overhead when fault-tolerant devices with active error correction are available.

%
%

\section*{Acknowledgements}
We thank Ken Xuan Wei for discussion about the CS gate calibration and providing us with the pulse sequence to investigate local coherent errors.
DCM and SS acknowledge partial support from the ARO under Contract No. W911NF-14-1-0124.

\bibliography{bibliography}

%
%

\appendix

\section{Basis Gate Information} \label{sec:device_info}

\begin{figure}[htbp]
  \includegraphics[width=0.95\linewidth]{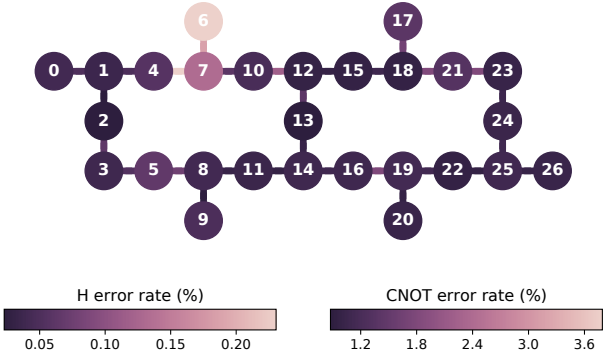}
  \caption[justified]{
  Distribution of single-qubit and two-qubit gate errors of \texttt{ibmq\_paris} at the time of experiment.
  Single-qubit gate errors measured by the Hadamard operation are shown in nodes of the qubit coupling map, while two-qubit gate errors measured by CNOT operation are shown in graph edges.
  Error values are represented by color maps shown in the bottom.
  }
  \label{fig:gate_error_map}
\end{figure}

In this paper all experiments are performed via cloud access to IBM Quantum system \texttt{ibmq\_paris}.
The backend provider calibrates single-qubit and two-qubit basis gates on a regular basis and provides pulse schedules and gate errors to users.
The gate error distribution at the time of experiment (2020-05-20 05:48 UTC) is shown in Fig. \ref{fig:gate_error_map}.
The averaged single-qubit gate error is $5.0 \times 10^{-4}$, while that of two-qubit gates is $1.4 \times 10^{-2}$.
The single-qubit gate error of the qubit 0 and 1, which are use in the CS gate, are $4.0 \times 10^{-4}$ and $3.7 \times 10^{-4}$, respectively.
The two-qubit CNOT gate error between these qubits is $1.3 \times 10^{-2}$.

\section{Quantum process tomography} \label{sec:qpt_experiment}

QPT was done using convex maximum likelihood estimation fitter with completely positive and trace-preserving (CPTP) constraints from the tomography module of Qiskit Ignis. The preparation basis $\{\ket{0}, \ket{1}, \ket{+}, \ket{+i}\}$ and measurement basis $\{X, Y, Z\}$ was used for each qubit.
We performed $M = 1024$ repetitions (shots) for each QPT basis configuration and readout error calibration circuit.
This requires 148 different experimental circuit executions per single $r_g^{\rm qpt}$ evaluation.
The readout error calibration circuit data was used to construct a 2-qubit measurement assignment matrix characterizing the Z-basis classical readout errors~\cite{Bravyi2020}.
This was used to compute noisy measurement basis POVM elements in the QPT fitter objective function to apply readout error mitigation during the QPT fit.
Note that this only mitigates the readout errors from the final Z-basis measurement.
Measurement errors arising from gate errors in the gates to change tomography measurement bases will not be affected.

The interval of each experiment trigger in this device is set to 1000 $\mu$s, therefore the minimum execution time of the whole experiment is estimated to be about 2.5 minutes.

\section{Calibrating CS Gate} \label{sec:gate_calibration}

\begin{figure}[htbp]
  \includegraphics[width=0.95\linewidth]{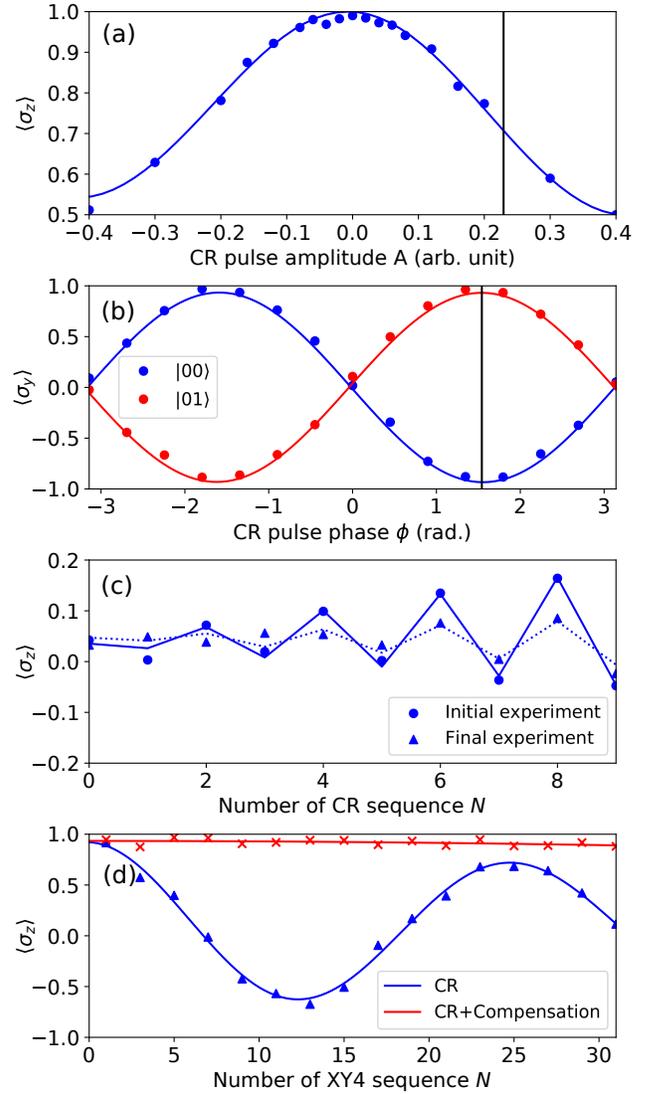}
  \caption[justified]{
  Typical experimental results for calibration experiments. Measured population is converted into the expectation value of Pauli operators.
  (a) Rough amplitude calibration.
  The blue and black line show the cosinusoidal fit for the experimental results and the optimal amplitude $A_0$.
  (b) Rough phase calibration.
  The blue and red line show the cosinusoidal fit for the experimental result of ${\cal S}_{\phi g}^{\rm scan}(\phi)$ and ${\cal S}_{\phi e}^{\rm scan}(\phi)$, respectively.
  The black line show the optimal phase $\phi_0$.
  (c) Rough amplitude calibration.
  The solid and dotted line show the fit for the result of initial ($A=0.236$) and final experiment ($A=0.237$) within the closed-loop calibration. The cosinusoidal function is used for the fit with $N$-dependent decay and baseline $F(N) = e^{-\alpha N} \cos(4(\pi/4 + \delta_A)N + \pi/2) + aN + b$.
  Here $\alpha, a$ and $b$ are additional fit parameters introduced empirically.
  The residual error per gate after the final experiment is $-1.25 \times 10^{-3}$ rad., which is lower than the threshold of $10^{-3}\pi$.
  (d) Compensation tone calibration.
  The blue and red line show the result of ${\cal S}^{\rm xy4}$ without and with the calibrated compensation tone, respectively.
  The cosinusoidal fit with decay for those curves yields crosstalk amplitude of 176.2 kHz and 6.7 kHz.
  All data in (a)--(c) are measured with $\tau_\text{sq} = 21.3$ ns, while (d) is measured with $\tau_\text{sq} = 0$ ns.
  }
  \label{fig:calibrations}
\end{figure}

The single qubit gates used for the echo sequence and local rotations are provided by \texttt{ibmq\_paris}.
We calibrate the CR pulse amplitude $A$ and its phase $\phi$ by the rough parameter scan followed by the closed-loop calibration.
These parameters are determined based on the two-pulse echoed CR sequence $U_{\rm echo}$ shown in Eq. \eqref{eq:echo}.
This approach simplifies the calibration, namely, we don't need to take non-negligible $ZI$ and $IX$ terms into account when we fit the experimental results for calibration parameters.
Calibrated sequence $U_{\rm echo} \sim [ZX]_\frac{\pi}{4}$ is used to realize the CS with local rotations shown in Eq. \eqref{eq:csgate}.

\subsection{Rough Parameter Scan}

We initialized both qubits in the ground state and perform a rough scan of the CR pulse amplitude with the pulse schedule:
\begin{align*}
  {\cal S}_{A}^{\rm scan}(A) &\equiv U_{\rm echo}(A, 0).
\end{align*}
The schedule is followed by the measurement of the target qubit in the $Z$-basis.
The sinusoidal fit for the measured population of the target qubit with ${\cal S}_A^{\rm scan}$ with different $A$ gives an estimate of the CR amplitude $A_0$ where the angle of controlled rotation is approximately $\pi/4$.
A typical experimental result for $\tau_{\rm sq} = 21.3$ ns is shown in Fig. \ref{fig:calibrations}(a).

By using this $A_0$, we scan the CR phase with two pulse schedules ${\cal S}_{\phi g}^{\rm scan}$ and ${\cal S}_{\phi e}^{\rm scan}$:
\begin{align*}
  {\cal S}_{\phi g}^{\rm scan}(\phi) &\equiv [IZ]_\frac{\pi}{2} \circ [IX]_\frac{\pi}{2} \circ U_{\rm echo}(A_0, \phi)^{2}, \\
  {\cal S}_{\phi e}^{\rm scan}(\phi) &\equiv [IZ]_\frac{\pi}{2} \circ [IX]_\frac{\pi}{2} \circ U_{\rm echo}(A_0, \phi)^{2} \circ [XI].
\end{align*}
The schedule ${\cal S}_{\phi g}^{\rm scan}$ (${\cal S}_{\phi e}^{\rm scan}$) drives the echo sequence $U_{\rm echo}(A_0, \phi)$ twice with the control qubit of the ground (excited) state.
Note that the last two operations correspond to the projection into $Y$-basis for the following measurement.
The flip of the state of the control qubit leads the controlled rotation of the target qubit state with opposite direction as illustrated in Fig. \ref{fig:calibrations}(b).
This opposite rotation of $\pi/2$ around an azimuthal angle $\theta = \theta_0 - \phi$ of the target qubit Bloch sphere yields measured outcome of $\mp$ 1 for ${\cal S}_{\phi g}^{\rm scan}$ and ${\cal S}_{\phi e}^{\rm scan}$, respectively, at the optimal phase $\phi = \phi_0$ where $\theta = 0$.
Here $\theta_0$ is the phase offset from the unknown transfer function of the coaxial cable assembly \cite{Krinner2019}.
The phase $\phi_0$ gives a rough estimate of the CR phase where the $ZX$ term of interest is maximized while the unwanted $ZY$ term is eliminated.

\subsection{Closed-loop Fine Calibration}

We use the roughly estimated parameters $(A_0, \phi_0)$ as an initial guess of closed-loop calibrations.
We first optimize the CR pulse amplitude with following experiment:
\begin{align*}
  {\cal S}_{A}^{\rm fine}(A) &\equiv U_{\rm echo}(A, \phi_0)^{4N} \circ [IX]_{\frac{\pi}{2}},
\end{align*}
where $N$ is number of repeated sequences.
This schedule prepares the target qubit in the superposition state and repeat the echo sequence $4N$ times to apply a controlled rotation of $N\pi$.
Because the initial guess of $A_0$ is estimated by the parameter scan in the coarse precision with a finite error $\delta_A$, repeating ${\cal S}_{A}^{\rm fine}$ for different $N$ can accumulate $\delta_A$ and this error appears as over rotation from the superposition state, as shown in Fig. \ref{fig:calibrations}(c).
The fit for the over rotation as a function of $N$ yields precise estimate of $\delta_A$, and we iteratively update the initial guess to optimize the CR pulse amplitude to $A_1$ where $\delta_A \sim 0$.
Here we use $N = 0, 1, 2 ..., 9$ and repeat updating the CR amplitude until the over rotation error reaches below the threshold value of $10^{-3}\pi$ rad.

With the optimized amplitude $A_1$, we tune the CR phase with following experiment:
\begin{align*}
  {\cal S}_{\phi}^{\rm fine}(\phi) &\equiv [IY]_{\frac{\pi}{2}} \circ (U_{\rm echo}(A_1, \phi) \circ [IY])^{N} \circ [IX]_{\frac{\pi}{2}}.
\end{align*}
This sequence also accumulates the small phase error $\delta_\phi$ as function of $N$.
We iteratively update the CR phase until the same threshold value with the amplitude calibration to obtain the optimal phase $\phi_1$ where $\delta_\phi \sim 0$.

\section{Crosstalk Estimation} \label{sec:xtalk_estimation}

The unwanted local rotation terms $IX$ and $IY$ can be simultaneously amplified with the following sequence combined with the XY-4 dynamical decoupling \cite{Ahmed2013} on the control qubit:
\begin{align*}
  {\cal S}^{\rm xy4} &\equiv ([YI] \circ U_{CR} \circ [XI] \circ U_{CR})^{2N},
\end{align*}
where $U_{CR} = U_{CR}(A_1, \phi_1)$.
Here, the CR pulse with the same sign is repeatedly applied while changing the state of control qubit.
This pulse sequence refocuses (and hence eliminates) controlled rotation terms such as $ZX$ and $ZY$, allowing us to precisely estimate the strength of weak local rotation terms $\sqrt{\omega_{IX}^2 + \omega_{IY}^2}$, amplified in the absence of strong two-qubit interactions.

This technique can be used to calibrate a compensation tone that eliminates the $IY$ term caused by the physical crosstalk between qubits \cite{Sheldon2016}.
The compensation tone is applied to the drive channel of the target qubit \texttt{d1}, in parallel with $U_{CR}$.
This single-qubit pulse is shaped as a flat-top pulse with Gaussian edges of identical duration as the $U_{CR}$ pulse, with its own calibrated amplitude and phase $(A', \phi')$.
First, we repeat ${\cal S}^{\rm xy4}$ for $N = 0, 2, 4, ..., 32$ without the compensation tone and measure the Pauli $Z$ expectation value of the target qubit.
The fit for the oscillation over the total CR gate time $8\tau_\text{CR}N$ yields the strength of the total unwanted local rotation terms.
At $\tau_{\rm sq} = 0$ ns, the unwanted local rotation strength of 176.2 kHz was observed.
This strength was reduced to 6.7 kHz with the calibrated compensation tone with $A' = 0.00102$ and $\phi' = -0.962$ rad.
The experimental result is shown in Fig. \ref{fig:calibrations}(d).

\end{document}